\documentclass[aip,reprint,citeautoscript]{revtex4-2}

\usepackage[colorlinks,urlcolor=blue,citecolor=blue]{hyperref}
\usepackage{color}
\usepackage{graphicx}
\usepackage{tabularx}
\usepackage{amsmath,amssymb}
\usepackage{multirow}

\usepackage[per-mode=symbol]{siunitx}
\DeclareSIUnit\angstrom{\text{\AA}}
\DeclareSIUnit\au{\text{at.\,u.}}

\usepackage[dvipsnames]{xcolor}

\begin{document}

\title{Reproducibility of real-time time-dependent density functional theory calculations of electronic stopping power in warm dense matter}

\author{Alina Kononov}
\email{akonono@sandia.gov}
\affiliation{Center for Computing Research, Sandia National Laboratories, Albuquerque NM 87123, USA \looseness=-1}

\author{Alexander J. White}
\email{alwhite@lanl.gov}
\affiliation{Los Alamos National Laboratory, Los Alamos, New Mexico 87545, USA \looseness=-1}

\author{Katarina A. Nichols}
\affiliation{Laboratory for Laser Energetics, University of Rochester, Rochester, New York 14623, USA \looseness=-1}
\affiliation{Department of Physics and Astronomy, University of Rochester, Rochester, New York 14623, USA \looseness=-1}

\author{S. X. Hu}
\affiliation{Laboratory for Laser Energetics, University of Rochester, Rochester, New York 14623, USA \looseness=-1}
\affiliation{Department of Physics and Astronomy, University of Rochester, Rochester, New York 14623, USA \looseness=-1}
\affiliation{Department of Mechanical Engineering, University of Rochester, Rochester, New York 14623, USA \looseness=-1}

\author{Andrew D. Baczewski}
\affiliation{Center for Computing Research, Sandia National Laboratories, Albuquerque NM 87123, USA \looseness=-1}

\begin{abstract}
Real-time time-dependent density functional theory (TDDFT) is widely considered to be the most accurate available method for calculating electronic stopping powers from first principles, but there have been relatively few assessments of the consistency of its predictions across different implementations.
This problem is particularly acute in the warm dense regime, where computational costs are high and experimental validation is rare and resource intensive. 
We report a comprehensive cross-verification of stopping power calculations in conditions relevant to inertial confinement fusion conducted using four different TDDFT implementations. 
We find excellent agreement among both the post-processed stopping powers and relevant time-resolved quantities for alpha particles in warm dense hydrogen.
We also analyze sensitivities to a wide range of methodological details, including the exchange-correlation model, pseudopotentials, initial conditions, observable from which the stopping power is extracted, averaging procedures, projectile trajectory, and finite-size effects.
We show that among these details, pseudopotentials, trajectory-dependence, and finite-size effects have the strongest influence, and we discuss different strategies for controlling the latter two considerations.
\end{abstract}

\maketitle

\section{Introduction}
\label{sec:intro}

Warm dense matter (WDM) is a challenging regime where plasma models falter and experiments remain extremely difficult.
Recent efforts increasingly leverage first-principles approaches based on density functional theory (DFT) to benchmark, inform, and improve more computationally efficient models for WDM and beyond \cite{grabowski2020review,baczewski2021predictions,fiedler2022deep,hentschel2023improving,nichols2023time,ward2023accelerating,stanek2023review}.
The credibility of the underlying first-principles data then underpins the predictive power of the more efficient models.
The scarcity of experimental avenues for model validation heightens the importance of building confidence in first-principles results by ensuring reproducibility, scrutinizing methodological choices, and quantifying uncertainties.

DFT calculations boast a high standard of reproducibility for static or equilibrium properties of materials \cite{lejaeghere2016reproducibility} despite differences among numerical implementations such as basis sets, pseudopotentials, and eigensolver algorithms.
Simulations of excited electron dynamics using real-time time-dependent DFT (TDDFT) involve additional choices including the time-stepping algorithm, perturbation properties, simulation length, and techniques for extracting observable information from time series data.
While some of these choices may be viewed as convergence parameters, high computational costs and non-trivial interactions among different options can make errors difficult to characterize.
Furthermore, compared to DFT, TDDFT has fewer practitioners and widely available full-featured implementations. 
Thus, a similar standard of reproducibility remains to be established.

Here, we consider TDDFT calculations of electronic stopping power, the rate at which a moving ion loses energy to electrons in a medium.
In inertial confinement fusion, this quantity controls an important contribution to the hot spot energy balance as fusion products re-deposit their kinetic energy to heat the fuel \cite{zylstra2019alpha-particle}.
Therefore, accurate stopping powers are important for correctly predicting ignition requirements \cite{hurricane2019approaching}.

In this work, we evaluate the reproducibility of TDDFT stopping power calculations and systematically examine the influence of different numerical and methodological choices.
While previous studies have validated and/or cross-verified final stopping power results \cite{ding2018ab, white2018time, zylstra2015measurement, kononov2021anomalous, malko2022proton}, those comparisons focus on a single post-processed quantity and leave open the possibility of fortuitous error cancellations that obscure the true degree of reproducibility.
Published comparisons of time-resolved TDDFT data \cite{andrade2021inq,kononov2023trajectory} remain rare but reveal significant discrepancies in the core contribution to stopping power that were attributed to details of the pseudopotentials \cite{kononov2023trajectory}.

We consider a simpler case without the sensitivities of core electrons: alpha-particle stopping power in warm dense hydrogen at a density of \SI{1}{\gram\per\centi\meter\cubed} and temperature of \SI{2}{\electronvolt}.
We further focus on a single alpha-particle velocity of \SI{5}{\au}, a choice that lies above the Bragg peak and involves both localized and delocalized excitations.
This test system enables rigorous assessment of pseudopotential effects by benchmarking against all-electron calculations with bare Coulomb potentials.
We compare both time-resolved data and average stopping powers computed using four different TDDFT codes: 
a custom extension \cite{baczewski2014numerical,magyar2016stopping} of the Vienna \emph{Ab initio} Simulation Package (\textsc{VASP}) \cite{kresse1996efficient,kresse1996efficiency,kresse1999from},
Stochastic and Hybrid Representation Electronic structure by Density functional theory (SHRED) \cite{Sharma2023,White_2022,White2020},
Qball \cite{schleife2012plane-wave,draeger2017massively},
and Inq \cite{andrade2021inq}.

We demonstrate good agreement across all TDDFT implementations, especially when matching methodological choices as much as possible.
Further analysis of sensitivities to different choices reveals which differences among computational approaches influence predicted stopping powers most.
This work clarifies and deepens understanding of best practices in TDDFT stopping power calculations and supports ongoing efforts to quantify uncertainties in large-scale simulations of high-energy-density systems \cite{stanek2023review}.

\section{Methods}
\label{sec:methods}

The central numerical task of a real-time TDDFT calculation \cite{ullrich2012time-dependent} consists of solving the time-dependent Kohn-Sham (TDKS) equations,
\begin{equation}
    i \frac{\partial}{\partial t} \phi_j(\mathbf{r},t) = H[n](\mathbf{r}, t) \, \phi_j(\mathbf{r},t) ,
    \label{eq:tdks}
\end{equation}
subject to some initial condition $\phi_j(\mathbf{r},t=0)$ and closure
\begin{equation}
    n(\mathbf{r},t) = \sum_j f_j(T_e) \left| \phi_j(\mathbf{r},t) \right|^2,
    \label{eq:den}
\end{equation}
where $\phi_j$ are Kohn-Sham (KS) electronic orbitals with temperature-dependent occupations $f_j(T_e)$ and $n(\mathbf{r}, t)$ is the electron density.
The Kohn-Sham Hamiltonian $H$ is a functional of the electron density and a sum of kinetic energy, external potential, Hartree, and exchange-correlation (XC) terms:
\begin{equation}
    H[n](\mathbf{r},t) = T + V_\mathrm{ext}(\mathbf{r},t) + V_\mathrm{Har}[n] + V_\mathrm{XC}[n],
    \label{eq:ham}
\end{equation}
where $V_\mathrm{ext}$ includes electron-ion interactions, $V_\mathrm{Har}$ captures the classical Coulombic electron-electron interaction, and $V_\mathrm{XC}$ contains quantum-mechanical corrections to the electron-electron interaction.
Within a TDDFT stopping power calculation, electrons respond to a moving particle that introduces explicit time-dependence within $V_\mathrm{ext}$.

In addition to the usual approximations, challenges, and limitations of DFT calculations --- a microscopic simulation cell, a finite basis representing $\phi_j$, the form of $V_\mathrm{XC}$, pseudopotentials entering $V_\mathrm{ext}$, and rigorous formalisms for computing physical observables --- real-time TDDFT simulations also choose an initial condition and perturbation form, discretize time, and access a finite temporal range.
In this work, all TDDFT calculations used a plane-wave basis with a cutoff energy of at least \SI{2000}{\electronvolt}, began from Mermin-Kohn-Sham equilibrium states \cite{mermin1965thermal}, and held ionic velocities fixed.
Other aspects of the simulations varied as described in the following subsections and summarized in Table \ref{tab:methods}.

\begin{table*}
    \centering
    \begin{tabular}{c|c|c|c|c|c|c|c|c|c|c|c}
        \multirow{2}{*}{code}  &  electronic &
            pseudo- & \multirow{2}{*}{XC} & initial  & \multirow{2}{*}{setup} & cutoff &
                                                                    time-   & \multirow{2}{*}{time step}                   & simulation &  force & stopping\\
                & orbitals & potentials &&    condition &&  energy &          stepper &                 & time     & analyzed & estimate\\\hline
        VASP         & KS
                & none     & LDA     & $\alpha \in$ IC    &  cubic & \SI{2000}{\electronvolt} & CN     & \SI{0.02}{\angstrom}$/v$\footnote{Slower projectiles with $v\leq\SI{3.5}{\au}$ required a smaller time step of \SI{0.25}{\atto\second}.} & \SI{80}{\angstrom}$/v$    & $F_\mathrm{tot}$         & $S_\mathrm{fit}$ \\
        SHRED                  & mixed
                & none      & PBE     & $\alpha \not\in$ IC  & elongated    & \SI{2000}{\electronvolt} &  SIL     &    0.121 as                      & \SI{31}{\angstrom}$/v \, \times \, 2$\footnote{``$\times\, 2$'' indicates that results were averaged over 2 trajectories.}                   & $\Delta F_\mathrm{ei}$ & $S_\mathrm{avg}$  \\
        SHRED & KS & HGH & PBE & $\alpha \not\in$ IC & elongated & \SI{4234}{\electronvolt} & SIL & \SI{0.048}{\atto\second} & \SI{0.27}{\femto\second} & $\Delta F_\mathrm{ei}$ & $S_\mathrm{avg}$ \\
        Qball   & KS & HSCV     & LDA     & $\alpha \in$ IC    & cubic  & \SI{2000}{\electronvolt} & ETRS   & \SI{0.02}{\angstrom}$/v$ & \SI{80}{\angstrom}$/v$     & $F_\mathrm{tot}$         & $S_\mathrm{fit}$ \\
        Inq     & KS & ONCV     & LDA     & $\alpha \in$ IC    & cubic  & \SI{2000}{\electronvolt} & ETRS   & \SI{0.02}{\angstrom}$/v$ & \SI{80}{\angstrom}$/v$    & $F_\mathrm{tot}$         & $S_\mathrm{fit}$
    \end{tabular}
    \caption{
    Differences among TDDFT simulations performed with each code to produce the data in Figs.\ \ref{fig:vdependence}.
    Throughout Sec.\ \ref{sec:results}, we consider several variations upon the choices listed here as described in the text.
    For the initial condition, $\alpha \in$ IC ($\alpha \not\in$ IC) indicates that the alpha particle projectile was included in (excluded from) from the Mermin-DFT calculation for the equilibrium electronic orbitals.
    In some cases, the time step and total simulation time varied inversely with the projectile velocity $v$.
    $F_\mathrm{tot}$ denotes the total force on the projectile and $\Delta F_\mathrm{ei}$ denotes the electron-ion contribution excluding the force on the projectile under a fixed initial electron density.
    $S_\mathrm{fit}$ and $S_\mathrm{avg}$ estimate average stopping power through a linear fit and direct averaging, respectively (see Eqs.\ \eqref{eq:work}\,--\,\eqref{eq:savg}).
    }
    \label{tab:methods}
\end{table*}

We obtain stopping powers from stopping forces $F(x) = -\mathbf{F}(x)\cdot \hat{v}$ against the projectiles's direction of motion $\hat{v}$ that depend implicitly on time through the distance $x(t)$ traveled by the alpha particle.
Different workflows rely on either the total force $F_\mathrm{tot}$ or the difference in the electron-ion contribution relative to a fixed electron density $\Delta F_\mathrm{ei}$ (see Table \ref{tab:methods}).
In either case, a time average of the stopping forces gives the average stopping power accessed by experiments and entering into larger-scale hydrodynamic simulations.

Again, different choices exist for estimating the average stopping power from time-dependent forces.
One method first computes the stopping work $W$ (i.e., energy deposited by the projectile) as a function of $x$ 
\begin{equation}
    W(x) = \int_{x_0} ^{x} F(x') \; dx'
    \label{eq:work}
\end{equation}
and then fits $W(x)$ to a linear model,
\begin{equation}
    W(x) \approx S_\mathrm{fit}(x_f) x + W_0 \qquad\mathrm{for} \;\; x_0<x<x_f, 
    \label{eq:sfit}
\end{equation}
where $S_\mathrm{fit}$ and $W_0$ are model parameters and $[x_0, x_f]$ is the range of projectile displacements included in the analysis.
Another approach averages the stopping force directly, effectively computing
\begin{equation}
    S_\mathrm{avg}(x_f) = \frac{W(x_f) - W(x_0)}{x_f-x_0}.
    \label{eq:savg}
\end{equation}

\begin{figure}
    \centering
    \begin{tabularx}{\columnwidth}{ll}
    (a) & \hspace{0.1in} (b) \\
    \includegraphics[height=1.25in]{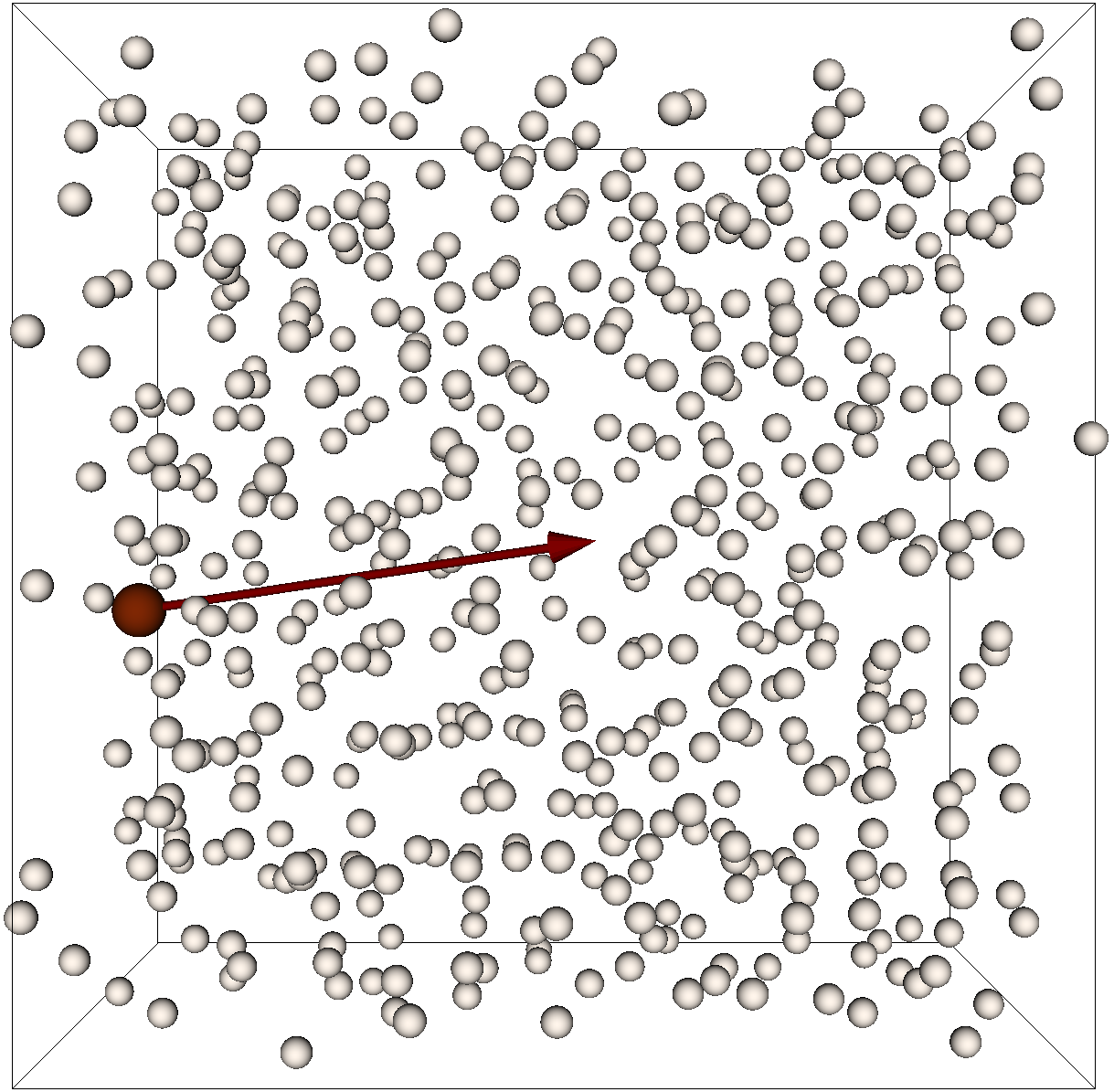} & \hspace{0.1in}
    \raisebox{0.15in}{\includegraphics[height=1in]{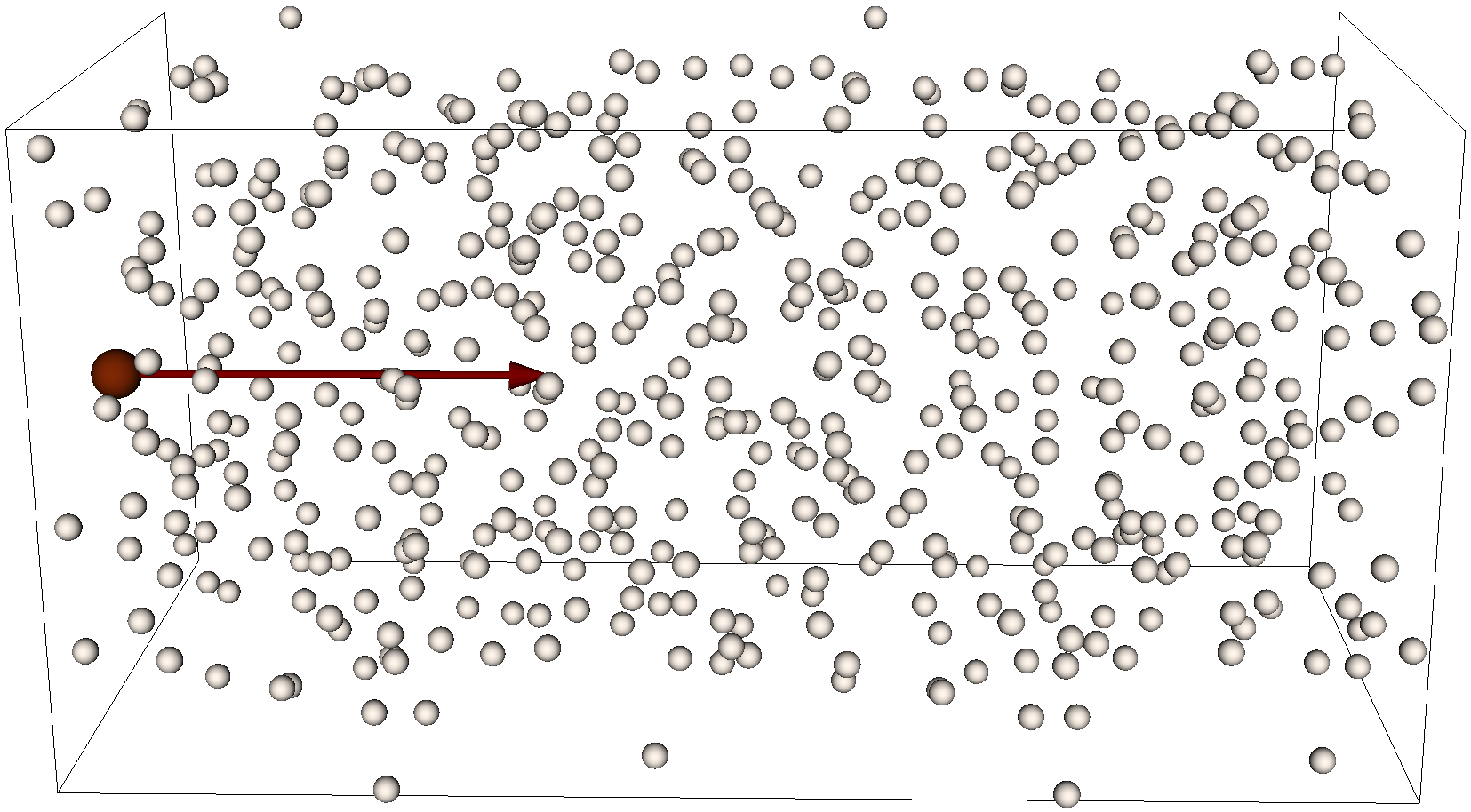}}
    \end{tabularx}
    \caption{
    (a) Cubic and (b) elongated simulation setups considered in this work.
    Both contain one alpha particle projectile (red) moving through a 512-atom hydrogen (white) configuration with trajectories indicated by red arrows.
    The hydrogen geometries were obtained from MD simulations as described in Appendix \ref{app:md}.
    In (a), the alpha particle's initial position and direction of motion were optimized according to the methods proposed by Ref.\ \onlinecite{kononov2023trajectory}.
    In (b), the alpha particle begins at an arbitrary point on the left edge of the supercell and travels in the long direction.
    }
    \label{fig:geometry}
\end{figure}

Calculations aimed at making direct comparisons of time-dependent data began from the same atomic configuration for the hydrogen host material and initial position and velocity for the alpha projectile.
We selected two setups reflecting different approaches commonly used by our groups: (i) a cubic simulation cell and an off-axis alpha particle trajectory optimized according to the methods developed by Kononov et al.\ \cite{kononov2023trajectory} (see Fig.\ \ref{fig:geometry}a) and (ii) an elongated simulation cell and the alpha particle travelling parallel to the long axis, following the approach of earlier work by Nichols et al. \cite{nichols2023time} (see Fig.\ \ref{fig:geometry}b).
For both setups, the atomic configuration was generated from separate \emph{ab initio} molecular dynamics (MD) simulations as described in Appendix \ref{app:md}. 
Our comparisons often focus on 512-atom hydrogen configurations, though we also consider other simulation cell sizes where indicated.

\subsection{VASP calculations}
\label{sec:vasptddft}

Although typical VASP \cite{kresse1996efficient,kresse1996efficiency,kresse1999from} calculations use the projector augmented-wave (PAW) method \cite{blochl1994projector}, the light elements involved in this work also allow all-electron calculations with no pseudopotentials.
By eliminating computations involving nonlocal projectors while only modestly increasing the required cutoff energy, bare Coulomb potentials significantly reduce computational costs for this system.
Thus, we used bare Coulomb potentials to maximize both accuracy and efficiency of VASP calculations, and we evaluate the influence of the pseudopotential approximation in Sec.\ \ref{sec:model}.

The adiabatic local density approximation (LDA) \cite{zangwill1980resonant,zangwill1981resonant} was used in VASP calculations throughout this work, and we compare VASP results for LDA and Perdew-Burke-Ernzerhof (PBE) \cite{perdew1996generalized} functionals in Sec.\ \ref{sec:model}.
About 1.3 electronic bands per hydrogen atom were needed to capture Fermi occupations of at least $10^{-5}$ at a temperature of \SI{2}{\electronvolt}.
For the $v=\SI{5}{\au}$ case scrutinized throughout this work, 
plane-wave cutoffs of \SI{2000}{\electronvolt} and \SI{2500}{\electronvolt} produced average stopping powers differing by less than 1\% and time-dependent forces differing by less than \SI{2.3}{\electronvolt\per\angstrom} after the \SI{4}{\angstrom} transient regime.

The TDDFT extension \cite{baczewski2014numerical,magyar2016stopping} of VASP uses the Crank-Nicolson (CN) time-stepping algorithm and solves the associated update equations using the Generalized Minimal Residual (GMRES) method~\cite{saad1986gmres} with a GMRES relative tolerance of $10^{-9}$.
For fast projectiles with $v>\SI{3.5}{\au}$, we set the time step inversely proportional to the projectile velocity such that the alpha particle traveled about \SI{0.02}{\angstrom} within each time step.
Slower projectiles required a shorter time step of \SI{0.25}{\atto\second} to adequately resolve the electron dynamics.
These choices converge average stopping power within 1\% (6\%) and time-dependent forces within \SI{1}{\electronvolt\per\AA} (\SI{5}{\electronvolt\per\angstrom}) relative to a time step half as long for an alpha particle with \SI{5}{\au} (\SI{1.5}{\au}) of velocity.

\subsection{SHRED calculations}
\label{sec:shredtddft}
In addition to the deterministic Kohn-Sham formulation used by the other codes, SHRED \cite{Sharma2023,White_2022,White2020} also includes capabilities for orbital-free, stochastic, and mixed stochastic-deterministic TDDFT.
In this work, we use both the deterministic and mixed stochastic-deterministic TDDFT implementations within SHRED.
The latter approach generates an initial finite-temperature Mermin density matrix from a subset of occupied KS orbitals and the complementary (orthogonal to the deterministic) stochastic density matrix.
The stochastic density matrix has a compressed low-rank form similar to the KS density matrix, thus stochastic ``orbitals'' are propagated in time following the same procedure as the KS orbitals.
The full procedure is described in Ref.\ \onlinecite{White_2022}.
For the 1024 (512) atom configuration we use 384 (192) deterministic orbitals and 64 stochastic ``orbitals'', with the stochastic representation accounting for about $25\%$ of the electrons. 
Comparisons to deterministic DFT with 640 KS orbitals at low (\SI{1.5}{\au}) and high (\SI{5}{\au}) velocities show a difference of less than 1\% percent.

The SHRED calculations used the PBE exchange-correlation functional \cite{perdew1996generalized} and either bare Coulomb potentials or Hartwigsen-Goedeker-Hutter (HGH) pseudopotentials \cite{goedecker1996separable,hartwigsen1998relativistic}, which have an analytic form that allows direct control of the cutoff radius.
The short iterative Lanczos (SIL) algorithm is used to propagate the Kohn Sham wavefunctions and stochastic vectors.\cite{Park86} 
The propagation is fully explicit with a time step of \SI{0.121}{\atto\second} or less.

Rather than computing the total force on the projectile as in the other TDDFT implementations, SHRED calculations typically aim to isolate force contributions that do not average to zero over time.
TDDFT studies commonly assume that the force induced by host ions $F_\mathrm{ii}$ has a zero time average because of symmetry along periodic trajectories or because nuclear stopping power is small in the velocity regime of interest.
Similarly, the force induced by the initial electron density matrix --- if it were held fixed as the projectile moved through it --- would also average to 0 due to the same symmetry.
This force, denoted as $F_\mathrm{ei}^{(0)}$, is expected to dominate instantaneous forces at high velocities, when the stopping power is small and the projectile only slightly perturbs the initial electron density.
Therefore, the SHRED implementation typically computes electron-ion force differences $\Delta F_\mathrm{ei} = F_\mathrm{ei} - F_\mathrm{ei}^{(0)}$.

\subsection{Qball calculations}
\label{sec:qball}
The Qball \cite{schleife2012plane-wave,draeger2017massively} calculations used Hamann-Schluter-Chiang-Vanderbilt (HSCV) norm-conserving pseudopotentials \cite{vanderbilt1985optimally} with outermost cutoff radii of \SI{0.8}{\au} (\SI{0.5}{\au}) for the alpha particle (hydrogen ions).
The electronic states evolved over time according to the enforced time-reversal symmetry (ETRS) algorithm \cite{castro2004propagators} with fourth-order Taylor expansions for the exponentials \cite{draeger2017massively}.
Rather than solving the nonlinear update equations to within a given tolerance, the ETRS implementation performs a two-step iteration that first estimates the electron density at the next time step, $n(\mathbf{r}, t+\Delta t)$, by evolving the KS states under $H(t)$.
The implicit portion of the propagator then uses the estimated $n(\mathbf{r}, t+\Delta t)$ to construct $H(t+\Delta t)$, with further self-consistent iteration not necessary for sufficiently small time steps \cite{draeger2017massively}.

Exchange and correlation was treated at the LDA \cite{zangwill1980resonant,zangwill1981resonant} level, and other numerical parameters were identical to those listed in Sec.\ \ref{sec:vasptddft} for VASP.
Independent convergence tests verified that the cutoff energy and time step also suffice for the different time stepper and pseudopotentials used within Qball.
For an alpha particle with \SI{5}{\au} of velocity, increasing the cutoff energy to \SI{2500}{\electronvolt} or halving the time step changes the average stopping power by less than 0.1\% and alters time-dependent forces by less than \SI{0.05}{\electronvolt\per\angstrom}.

\subsection{Inq calculations}
\label{sec:inq}
The Inq\cite{andrade2021inq} simulations follow very similar methodology as described for Qball in Sec.\ \ref{sec:qball}.
However, we used the default pseudopotential set within Inq: optimized norm-conserving Vanderbilt (ONCV) pseudopotentials \cite{hamann2013optimized} with outermost cutoff radii of \SI{1.25}{\au} (\SI{1.0}{\au}) for the alpha particle (hydrogen ions).
Since these pseudopotentials are even softer than those used in the Qball calculations, we expect excellent convergence at the same cutoff energy.
Also, the Inq ETRS implementation performs up to 5 self-consistent iterations to obtain $n(\mathbf{r}, t+\Delta t)$ and $H(t+\Delta t)$, further enhancing convergence with respect to time step.

\section{Results and Discussion}
\label{sec:results}
\begin{figure}
    \centering
    \includegraphics{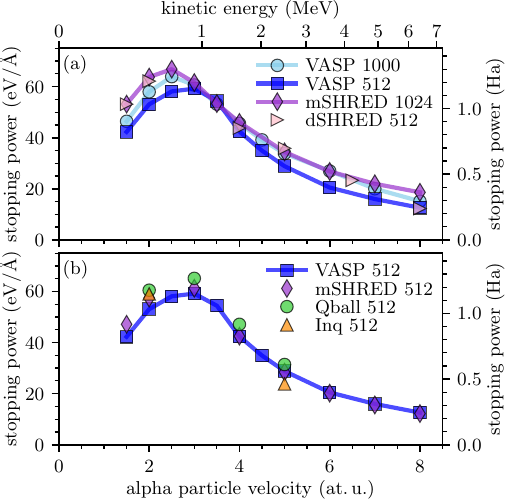}
    \caption{Average stopping powers in warm dense hydrogen (\SI{1}{\gram\per\centi\meter\cubed}, \SI{2}{\electronvolt}) as a function of the alpha particle velocity.
    Results are labelled by the TDDFT code (with corresponding methods summarized in Table \ref{tab:methods}) and the number of hydrogen atoms in the simulation.
    mSHRED and dSHRED denote mixed stochastic-deterministic and deterministic (KS) TDDFT variants within SHRED.
    In (a), the VASP (SHRED) calculations used cubic (elongated) simulations cells with off-axis (on-axis) trajectories.
    In (b), all simulations used the same cubic / off-axis setup.
    }
    \label{fig:vdependence}
\end{figure}

We present the average stopping powers computed by different TDDFT codes and methodologies (see Table \ref{tab:methods}) in Fig.\ \ref{fig:vdependence}.
Overall, we find reasonable agreement, though some of the discrepancies exceed the stringent numerical convergence achieved within each set of calculations (see Sec.\ \ref{sec:methods}).
In particular, we see excellent agreement between VASP and SHRED results when both sets of calculations use bare Coulomb potentials and the same 512-atom simulation setup (blue squares and purple diamonds in Fig.\ \ref{fig:vdependence}b), but modest deviations arise when using pseudopotentials and/or changing the simulation cell shape and alpha particle trajectory.

\begin{figure}
    \centering
    \includegraphics{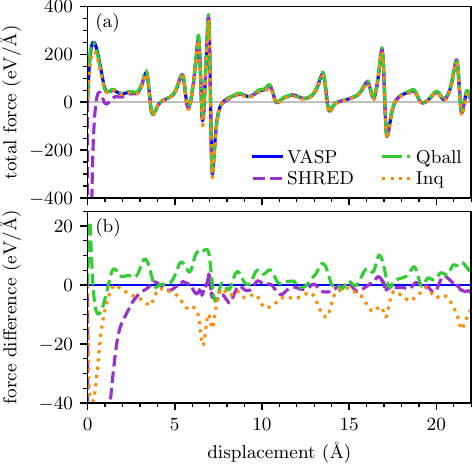}
    \caption{
    Total stopping forces on an alpha particle moving through warm dense hydrogen (\SI{1}{\gram\per\centi\meter\cubed}, \SI{2}{\electronvolt}) with 5 atomic units (at.\,u.) of velocity as computed by each TDDFT code for the 512-atom cubic test system.
    (a) shows the force against the direction of projectile motion and (b) shows differences from the VASP data (solid blue).
}
    \label{fig:allcodes}
\end{figure}

Since stopping power is a highly integrated quantity, we also compare the instantaneous friction force experienced by the projectile along its path for the case of $v=\SI{5}{\au}$ in \mbox{Fig.\ \ref{fig:allcodes}}, where all calculations used the same off-axis alpha particle trajectory through the same 512-atom hydrogen configuration within the cubic simulation setup (see Fig.\ \ref{fig:geometry}a), i.e., the same setup used for Fig.\ \ref{fig:vdependence}b.
Overall, we find very good agreement in these stopping forces, particularly after an initial transient regime during the first few \AA.
However, small relative differences in the stopping forces accumulate into noticeable differences in average stopping power, which comes out to 
\SI{28.8}{\electronvolt\per\angstrom} from VASP, 
\SI{31.4}{\electronvolt\per\angstrom} from Qball, 
\SI{30.9}{\electronvolt\per\angstrom} from SHRED, and 
\SI{23.8}{\electronvolt\per\angstrom} from Inq
when using the 512-atom cubic setup, identical post-processing methods to extract $S_\mathrm{fit}$ from $F_\mathrm{tot}$, and other choices as listed in Table \ref{tab:methods}.

In the following subsections, we examine the extent to which different methodological variations contribute to the modest discrepancies occurring in \mbox{Figs.\ \ref{fig:vdependence} and \ref{fig:allcodes}}.
Among the different choices listed in Table \ref{tab:methods}, we show in Sec.\ \ref{sec:model} that pseudopotentials can have a large and persistent effect on time-dependent stopping forces.
Meanwhile, the XC functional (LDA or PBE) and the initial condition (including or excluding the alpha-particle projectile) primarily alter forces during the first few \AA\ traversed and do not significantly affect average stopping powers for this system (see Secs.\ \ref{sec:model} and \ref{sec:ic}).
Sections \ref{sec:observable} and \ref{sec:averaging} demonstrate close agreement between average stopping powers estimated from different possible force observables and post-processing techniques.
Finally, Sec.\ \ref{sec:setup} reveals contrasting behavior of finite-size errors within the cubic and elongated simulation setups.

\subsection{Influence of model approximations}
\label{sec:model}
We begin by considering differences among the methods described in Sec.\ \ref{sec:methods} that alter the underlying Hamiltonian (Eq.\ \eqref{eq:ham}): the electron-ion potentials determining $V_\mathrm{ext}$ and the exchange-correlation (XC) functional $V_\mathrm{XC}$.
We expect the XC functional to have only a minor influence because LDA and PBE stopping powers differed by less than 0.5\% in an earlier study of a metallic system \cite{kononov2023trajectory}.
On the other hand, it is well known that pseudopotentials can cause severe underestimation of high-velocity stopping powers by neglecting contributions from pseudized core electrons \cite{schleife2015accurate}.
Recent work also showed that softer pseudopotentials limit energy transfer during close collisions with host nuclei, reducing stopping power predictions by around 10\% even for the same number of explicitly simulated electrons \cite{kononov2023trajectory}.
Although our light-ion test system contains no core electrons and all the calculations treated all of the electrons explicitly, pseudopotentials can still influence processes involving electrons localized around the nuclei.

\begin{figure}
    \centering
    \includegraphics{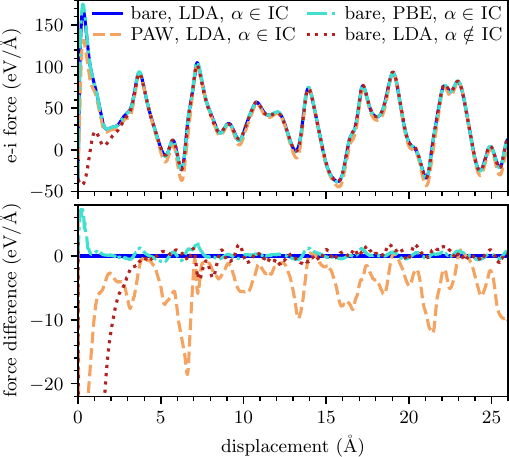}
    \caption{
    Electron-ion stopping forces on an alpha particle moving through warm dense hydrogen (\SI{1}{\gram\per\centi\meter\cubed}, \SI{2}{\electronvolt}) with \SI{5}{\au} of velocity for the cubic test system.
    We compare results from a series of VASP calculations using bare Coulomb or PAW potentials, LDA or PBE, and the alpha particle included in the initial condition ($\alpha\in$ IC) or excluded from it ($\alpha\notin$ IC).
    (a) shows the force against the direction of projectile motion and (b) shows differences from the case with bare Coulomb potentials, LDA, and the alpha particle included in the intial condition (solid blue).
    }
    \label{fig:vasp}
\end{figure}

In Fig.\ \ref{fig:vasp}, we compare VASP predictions obtained with LDA and PBE XC functionals and using bare Coulomb or PAW potentials for both the projectile and host ions.
We examine only the electron-ion contribution to the stopping forces because the ion-ion contribution dominating total forces remains identical across all cases. 
We find that LDA and PBE indeed produce very similar forces along the alpha particle's direction of motion, differing by at most \SI{2.4}{\electronvolt\per\angstrom} after the alpha travels \SI{1}{\angstrom} within VASP.
Although these deviations are significant relative to the average stopping power of \SI{28.8}{\electronvolt/\angstrom} for this case, their time-averages amount to less than 0.5\% of the average stopping power.
Therefore, we conclude that the choice of XC functional does not appreciably contribute to discrepancies among different codes in this work.

On the other hand, pseudopotentials can have a much stronger influence on stopping power: VASP calculations using PAW predict systematically weaker stopping forces than obtained with bare Coulomb potentials.
These deviations persist throughout the entire simulation length and lead to a 13\% lower average stopping power with PAW than with bare Coulomb potentials.
However, contrary to earlier findings for proton stopping in ambient aluminum \cite{kononov2023trajectory}, we find less than 0.1\% sensitivity to the hardness of the hydrogen PAW, suggesting that the projectile pseudopotential has a stronger influence in this case.

Similarly, different norm-conserving pseudopotentials within Qball or Inq produce significantly different stopping powers, either overestimating or underestimating by up to 18\% relative to VASP with bare Coulomb potentials.
Therefore, we attribute the larger deviations of Qball and Inq data from other codes in Figs.\ \ref{fig:vdependence} and \ref{fig:allcodes} to the pseudopotential approximation.
Surprisingly, stopping powers predicted from SHRED appear less sensitive to pseudopotentials, with only a 2\% difference between results computed using bare Coulomb potentials and HGH pseudopotentials with cutoff radii of \SI{1}{\au} for both species.

\subsection{Influence of initial condition}
\label{sec:ic}

\begin{figure}
    \centering
    \includegraphics{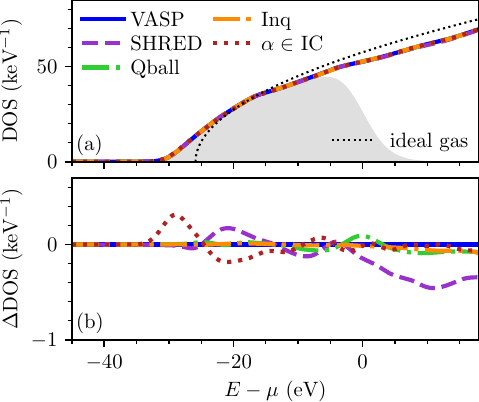}
    \caption{
    (a) Densities of states (DsOS) for hydrogen at \SI{1}{\gram\per\centi\meter\cubed} and \SI{2}{\electronvolt} computed using each code with the pseudopotentials and XC functionals listed in Table \ref{tab:methods}.
    To obtain smooth curves, these Mermin-DFT calculations used a denser $2^3$ $\Gamma$-centered reciprocal-space quadrature and the DsOS were obtained by applying a Gaussian broadening of \SI{2}{\electronvolt} to the Mermin-KS eigenvalues.
    The ``$\alpha\in$ IC'' curve (dotted red) was computed with VASP and included the alpha particle projectile.
    The DFT data departs modestly from the ideal Fermi gas (dotted black) at the corresponding electron density.
    Gray shading indicates the occupied DOS.
    In (b), differences are taken relative to the VASP DOS (solid blue).
    }
    \label{fig:dos}
\end{figure}

In addition to the form of the Hamiltonian, the solution to the TDKS equations (Eqs.\ \eqref{eq:tdks}\,--\,\eqref{eq:den}) also depends on the initial KS orbitals $\phi_j(\mathbf{r}, t=0)$.
In particular, the density of states (DOS) corresponding to the initial orbitals relates to the distribution of electronic transitions that the projectile could excite.
For this reason, the DOS enters into some stopping power models based on the average-atom method \cite{hentschel2023improving}, and close agreement for the initial DOS helps ensure agreement among stopping powers.

All the TDDFT simulations began from Mermin-KS equilibrium states \cite{mermin1965thermal}, and Fig.\ \ref{fig:dos} demonstrates that the Mermin-DFT implementations predict very similar electronic properties for the static, unperturbed hydrogen host even under different pseudopotentials and XC functionals (see Table \ref{tab:methods}).
The largest deviations among the DsOS in Fig.\ \ref{fig:dos} amount to less than 1\%.
Furthermore, including the alpha particle within the Mermin-DFT calculation has only a minor influence on the DOS of the initial system, and we expect that this effect will further diminish with increasing system size.

Although we find excellent agreement for initial electronic properties, different treatments of the projectile within the initial condition determine initial screening of the alpha particle and thus influence stopping forces during the TDDFT simulations.
The SHRED calculations exclude the projectile from the initial condition and insert it at $t=0$ within the TDDFT calculation, thereby simulating an initially fully ionized alpha particle.
On the other hand, VASP calculations typically include the projectile in the initial condition, resulting in an initially partially neutralized ion.
Previous Qball studies have shown that the projectile's initial charge state can affect results for high-Z projectiles \cite{lee2020multiscale} and pre-equilibrium behavior in 2D materials \cite{vazquez2021electron}, but light-ion stopping powers in bulk materials do not depend strongly on the initial condition \cite{schleife2015accurate} because the projectile's charge state quickly equilibrates after traversing a few \AA s \cite{kononov2020pre-equilibrium}.

Here, we revisit the sensitivity of stopping powers to the initial condition in the warm dense regime.
Consistent with earlier results for ambient materials, we find that the VASP stopping forces do not depend strongly on whether the alpha particle was included in the initial condition (see Fig.\ \ref{fig:vasp}).
After a transient regime within the first \SI{4}{\angstrom} of the alpha particle's path --- during which the ion dynamically captures or sheds electrons to reach an equilibrium charge state --- the initial condition changes stopping forces by at most \SI{3.8}{\electronvolt\per\angstrom}.
On average, these deviations again amount to less than 0.5\% of the average stopping power and thus do not meaningfully affect the code comparisons in this work.

\subsection{Influence of observable choice}
\label{sec:observable}

Even with the same underlying electron dynamics, it is possible to extract an average stopping power from different observables.
For example, Ref.\ \onlinecite{schleife2015accurate} previously demonstrated the equivalence of analyzing the total energy under fixed ion velocities and analyzing the total force on the projectile; these observables are related through the work done by maintaining a constant projectile velocity.
Here, we consider the role of different contributions to the total force on the projectile, 
\begin{equation}
    F_\mathrm{tot} = F_\mathrm{ei} + F_\mathrm{ii},
\end{equation}
where $F_\mathrm{ei}$ is the electron-ion force and $F_\mathrm{ii}$ is the ion-ion force.
We also consider 
\begin{equation}
\Delta F_\mathrm{ei} = F_\mathrm{ei} - F_\mathrm{ei}^{(0)},
\end{equation}
where $F_\mathrm{ei}^{(0)}$ represents the force that a fixed initial electronic system would induce, a contribution often excluded from SHRED calculations as described in Sec.\ \ref{sec:shredtddft}. 

\begin{figure}
    \centering
    \includegraphics{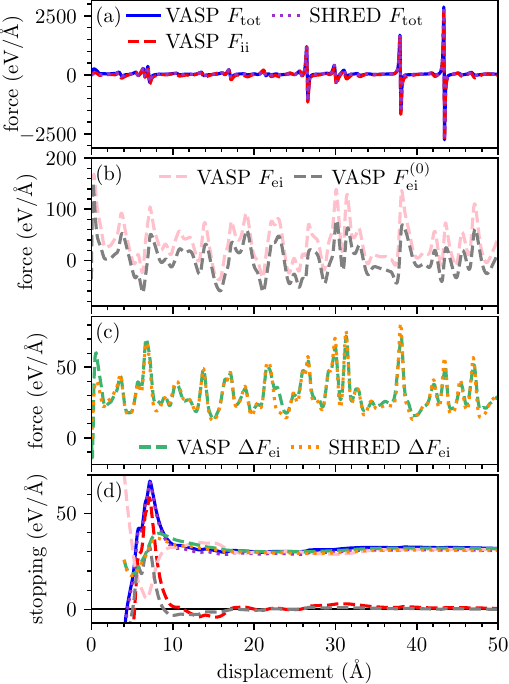}
    \caption{
    Different contributions to the stopping forces on an alpha particle moving through warm dense hydrogen (\SI{1}{\gram\per\centi\meter\cubed}, \SI{2}{\electronvolt}) with \SI{5}{\au} of velocity as computed by VASP and SHRED for the 512-atom cubic test system.
    Panels (a)\,--\,(c) compare the total force $F_\mathrm{tot}$, ion-ion force $F_\mathrm{ii}$, electron-ion force $F_\mathrm{ei}$, and the difference between electron-ion forces under time-evolving and static electron densities $\Delta F_\mathrm{ei} = F_\mathrm{ei} - F_\mathrm{ei}^{(0)}$.
    Panel (d) shows the average stopping power generated by each force contribution with colors corresponding to panels (a)\,--\,(c).
    }
    \label{fig:observable}
\end{figure}

In Fig.\ \ref{fig:observable}, we show that although the magnitude and time-dependent behavior of each force contribution exhibit stark differences, stopping powers inferred from $F_\mathrm{tot}$, $F_\mathrm{ei}$, and $\Delta F_\mathrm{ei}$ agree closely.
Although the ion-ion contribution $F_\mathrm{ii}$ dominates the absolute energy scale of the total force (cf.\ Figs.\ \ref{fig:observable}a and b), its time average quickly decays toward $0$ and amounts to 2.6\% of the converged total stopping power (see Fig.\ \ref{fig:observable}d).
Similarly, including the force due to the unperturbed electron density $F_\mathrm{ei}^{(0)}$ changes the extracted stopping power by only 0.6\% within VASP and 0.2\% within SHRED.

Contributions from $F_\mathrm{ii}$ and $F_\mathrm{ei}^{(0)}$ do not depend on projectile velocity, but become more significant relative to lower stopping powers in the high-velocity regime.
For the fastest alpha particle considered in this work, $v=\SI{8}{\au}$, including $F_\mathrm{ii}$ and $F_\mathrm{ei}^{(0)}$ changes the stopping power by 6\% and 1.4\%, respectively.
Thus, excluding these terms becomes more important for precise calculations of electronic stopping power at high velocities.

\subsection{Influence of averaging procedure}
\label{sec:averaging}

Given a choice of observable, different analysis methods are possible for estimating an average stopping power (see Eqs.\ \eqref{eq:sfit} and \eqref{eq:savg}).
Ref.\ \onlinecite{kononov2023trajectory} previously showed that when analyzing the total force, $S_\mathrm{fit}$ behaves more smoothly and converges faster over time than $S_\mathrm{avg}$.
Here, we revisit the behavior of both methods when analyzing $F_\mathrm{ei}$ and $\Delta F_\mathrm{ei}$.

\begin{figure}
    \centering
    \includegraphics{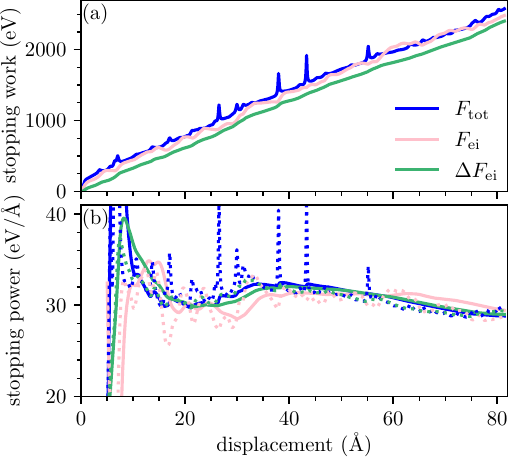}
    \caption{
    (a) Stopping work $W(x)$ (see Eq.\ \eqref{eq:work}) as computed from different forces within VASP on an alpha particle moving through warm dense hydrogen (\SI{1}{\gram\per\centi\meter\cubed}, \SI{2}{\electronvolt}) with \SI{5}{\au} of velocity.
    (b) Average stopping power estimated as $S_\mathrm{fit}$ (solid, see Eq.\ \eqref{eq:sfit}) and $S_\mathrm{avg}$ (dotted, see Eq.\ \eqref{eq:savg}).
    }
    \label{fig:analysis}
\end{figure}

In Fig.\ \ref{fig:analysis}a, the stopping work computed from $F_\mathrm{tot}$ contains sharp features arising from strong occasional ion-ion force pulses (see Fig.\ \ref{fig:observable}a).
Consistent with Ref.\ \onlinecite{kononov2023trajectory}, similar sharp features appear in $S_\mathrm{avg}$ because this estimate depends upon only $W(x_f)$ and $W(x_0)$, whereas $S_\mathrm{fit}$ provides a smoother estimate because it's informed by a range of $W(x)$ data (see Fig.\ \ref{fig:analysis}b).
Excluding ion-ion forces and/or the force induced by the equilibrium density --- i.e., analyzing $F_\mathrm{ei}$ or $\Delta F_\mathrm{ei}$ --- dramatically improves the behavior of $S_\mathrm{avg}$, but the qualitative behavior of $S_\mathrm{fit}$ is less sensitive to the choice of force and still maintains somewhat smoother evolution in all cases.
All combinations of stopping force and averaging method agree within 4\% by a projectile displacement of \SI{80}{\angstrom}, though these choices would contribute to uncertainties for shorter simulation lengths and may become more important for higher-Z host materials with core electrons contributing sharp features to $F_\mathrm{ei}$ and/or higher projectile velocities where $F_\mathrm{ei}^{(0)} \gg \Delta F_\mathrm{ei}$.

\subsection{Influence of simulation setup}
\label{sec:setup}

Finally, we compare data obtained from the two different simulation setups: cubic supercell with off-axis trajectory and elongated supercell with on-axis trajectory.
The two approaches differ in terms of the environments sampled by the alpha particle and the behavior of finite-size errors, and these factors influence how well the computed stopping powers reflect practically meaningful values independently of the approximations and choices discussed above.
To disentangle the two effects, we consider two types of elongated supercells: 256\,--\,1024-atom cells constructed by replicating a thermalized 256-atom configuration and a fully thermalized 1024-atom cell.
The replicated cells provide information about finite-size errors under fixed environment sampling, while the thermalized cell represents a more converged production setup.

Naively, stopping powers in a hydrogen plasma might not be expected to depend strongly on environment sampling because the density of states resembles that of a Fermi gas (see Fig.\ \ref{fig:dos}).
Sampling considerations become important in systems and regimes where core electrons (or electrons with strong spatial nonuniformities) significantly contribute to stopping powers.
For example, earlier work \cite{schleife2015accurate,maliyov2020quantitative,kononov2023trajectory} found that obtaining accurate stopping powers beyond the Bragg peak in crystalline aluminum requires that the projectile representatively sample close collisions with host nuclei and the associated $2s$ and $2p$ orbital excitations.
In that system, no single on-axis trajectory achieves sufficiently representative sampling and only appropriately chosen off-axis trajectories produce stopping powers in agreement with empirical data.

\begin{figure}
    \centering
    \includegraphics{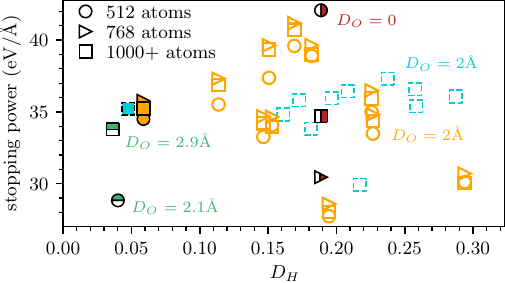}
    \caption{
    Variations in VASP stopping power predictions for different projectile trajectories and supercell configurations for warm dense hydrogen and an alpha particle velocity of \SI{5}{\au}. 
    $D_H$ quantifies deviations from ideal environment sampling, and $D_O$ is the minimum distance between the projectile and periodic images of its previously traversed path.
    Symbol shapes indicate supercell size, while colors and fill styles correspond to different schemes for obtaining an average stopping power.
    Empty symbols indicate random on-axis trajectories with a single pass through an elongated simulation cell such that the projectile stops \SI{2}{\angstrom} before reaching its starting point.
    Data points with a solid orange outline used simulation cells comprised of a 256-atom cell replicated in the direction of projectile motion, while dashed turquoise outlines used a fully equilibrated 1024-atom cell.
    Filled orange and turquoise points indicate aggregate results for corresponding trajectory ensembles.
    Semi-filled points use a single, \SI{80}{\angstrom}-long trajectory in the on-axis, elongated setup (red, right-filled) or the optimized off-axis, cubic setup (green, top-filled).
    }
    \label{fig:dh}
\end{figure}

Despite the nearly ideal DOS and lack of core electrons, the equilibrium electron density of the warm dense hydrogen host contains spatial variations spanning more than two orders of magnitude.
This nonuniformity causes surprisingly large sensitivities to the alpha particle trajectory.
Stopping powers computed using ten random on-axis trajectories through elongated supercells, i.e., using different initial positions for an alpha particle traveling in the long direction, deviated from their average by as much as 20\% (see Fig.\ \ref{fig:dh}).
While this trajectory-dependence appears weaker than in the aluminum case \cite{schleife2015accurate,maliyov2020quantitative,kononov2023trajectory}, it remains a significant source of uncertainty in these TDDFT calculations.

The length of the simulation cell fundamentally limits the sampling fidelity achievable by a single on-axis trajectory, as the environment would become periodic after one pass through the cell.
In contrast, an off-axis trajectory can continue sampling new environments and improving the sampling fidelity during each pass through the periodic cell while still avoiding its own wake \cite{kononov2023trajectory}.
Using an ensemble of several trajectories (either on-axis or off-axis) can further improve sampling fidelity \cite{gu2020efficient}.

Using the quantitative metric proposed by Ref.\ \onlinecite{kononov2023trajectory} to assess trajectory quality, the ten random on-axis trajectories deviate from ideal sampling by $D_H$ between 0.1 and 0.3 when considered individually.
When taken in aggregate, the ensemble of on-axis trajectories achieves a much lower $D_H$ value of 0.05\,--\,0.06, comparable to the sampling fidelity of the optimized off-axis trajectories in the cubic setup which achieve $D_H$ values around 0.04.
Interestingly, the distribution of $D_H$ values in the fully thermalized 1024-atom supercell (turquoise squares in \mbox{Fig.\ \ref{fig:dh}}) appears similar to the replicated supercells (orange squares) despite the less periodic environment sampled in the former case.
For the largest supercells considered in this work, stopping powers predicted from the optimized off-axis trajectory and an ensemble average over on-axis trajectories agree within 4\%, confirming the validity of both approaches.

The two simulation setups also differ in the behavior of finite-size errors, which TDDFT stopping power studies often attribute to the neglect of long-wavelength plasmonic excitations \cite{correa2018calculating}.
The elongated simulation cell supports longer wavelength plasmons along the projectile's direction of motion, reducing this contribution to finite-size error.
However, as the projectile crosses periodic boundary conditions, artificial interactions with earlier excitations cause significant deviations from the plasmonic model of finite-size errors \cite{kononov2023trajectory}.
These so-called ``ouroboros effects'' would be more severe for a long on-axis trajectory, where the projectile traverses the same path during each pass through the simulation cell.

Indeed, Fig.\ \ref{fig:dh} shows only modest (up to 5\%) stopping power sensitivity to supercell size for short on-axis trajectories that complete less than a single pass through the elongated cell (orange symbols), both for individual trajectories (empty symbols) and their ensemble average (filled symbols).
We note that this observation only holds for supercells long enough that each trajectory begins to approach a steady-state stopping power within the single pass; we find much larger errors of up to 50\% (23\%) for individual trajectories (ensemble average) in a shorter, 256-atom supercell.
In contrast, using a long on-axis trajectory (red right-filled symbols), which achieves steady-state stopping at the expense of making multiple passes over the same path through the elongated cell, incurs large and unsystematic finite-size errors of about 20\% even for cell sizes that are well-converged over a single pass.
We also find significant finite-size errors in the cubic setup with an optimized off-axis trajectory (green top-filled symbols), with a 15\% difference between 512-atom and 1000-atom stopping powers.

Overall, we find good agreement within about 4\% when appropriately controlling both environment sampling and finite-size errors using different approaches.
Depending on the approach, both effects can significantly influence the accuracy of stopping power predictions.
Long on-axis trajectories suffer from both strong finite-size effects and limited sampling fidelity.
While using short on-axis trajectories through elongated supercells converges very quickly with respect to simulation cell size, achieving representative environment sampling with this method requires an ensemble of trajectories.
Conversely, long off-axis trajectories can be optimized to achieve excellent sampling fidelity, but require larger supercells to converge finite-size errors.
Future work may investigate the ability of a compromising approach relying on intermediate-length off-axis trajectories to better balance computational costs in reducing both types of errors. 

\section{Conclusions}

Through detailed comparisons across four independent real-time TDDFT implementations, this work establishes reproducibility standards and quantifies uncertainties arising from different possible methodological choices in first-principles calculations of electronic stopping power.
For the case of alpha particles in warm dense hydrogen, the choice of XC functional (adiabatic LDA or PBE) and initial condition (including or excluding the projectile in the Mermin-KS equilibrium states) only negligibly influence average stopping power because these options only weakly alter underlying time-dependent forces beyond a short transient regime.
Similarly, for sufficiently long simulation times we find close agreement among different post-processing techniques involving various possible force contributions.

The dominant sources of error in TDDFT stopping powers calculations typically arise from pseudopotentials, poor sampling of the environment by the projectile, and/or finite simulation cells.
Even in the all-electron simulations considered here, pseudopotentials can systematically underestimate stopping powers by 13\% compared to bare Coulomb potentials.
The complex nature of environment sampling and finite-size errors has inspired different mitigation strategies, and direct comparison of their convergence behavior suggests that both effects remain significant even within this relatively simple test system.
Infrared truncation of plasmonic excitations, limited sampling of different plasma environments, and artificial interaction with the projectile's wake would still limit accuracy even if exact simulation of the many-body electron dynamics \cite{rubin2023quantum} became feasible, as obtaining experimentally-relevant values requires expensive averages over multiple trajectories, large supercells containing thousands of particles, and/or complicated techniques for optimizing trajectories.

\begin{acknowledgements}

We thank the organizers of the 2nd Charged-Particle Transport Coefficient
Comparison Workshop --- including Lucas J. Stanek, Brian M. Haines, Stephanie B. Hansen, Patrick F. Knapp, Michael S. Murillo, Liam G. Stanton, and Heather D. Whitley --- for catalyzing this work.
We are also grateful to 
Yifan Yao for technical advice,      
Joshua P.\ Townsend for sharing equilibrium H configurations, 
and Heath Hanshaw for pre-publication review.

AK and ADB were partially supported by the US Department of Energy Science Campaign 1 and Sandia National Laboratories' Laboratory Directed Research and Development (LDRD) Project No.\ 233196. 
AJW was supported by the LDRD program of Los Alamos National Laboratory (LANL), under Project Numbers 20210233ER and 20230323ER, and U.S. Department of Energy Science Campaign 4. This research used computing resources provided by the LANL Institutional Computing and Advanced Scientific Computing programs. Los Alamos National Laboratory is operated by Triad National Security, LLC, for the National Nuclear Security Administration of U.S. Department of Energy (Contract No. 89233218CNA000001).
KAN and SXH were supported by the Department of Energy [National Nuclear Security Administration] University of Rochester ``National Inertial Confinement Program'' under Award Number DE-NA0004144. 

This article has been co-authored by employees of National Technology \& Engineering Solutions of Sandia, LLC under Contract No. DE-NA0003525 with the U.S. Department of Energy (DOE). The authors own all right, title and interest in and to the article and are solely responsible for its contents. The United States Government retains and the publisher, by accepting the article for publication, acknowledges that the United States Government retains a non-exclusive, paid-up, irrevocable, world-wide license to publish or reproduce the published form of this article or allow others to do so, for United States Government purposes. The DOE will provide public access to these results of federally sponsored research in accordance with the DOE Public Access Plan \url{https://www.energy.gov/downloads/doe-public-access-plan}.

\end{acknowledgements}

\appendix
\section{Molecular dynamics simulations}
\label{app:md}

In the main text, we consider cubic and elongated simulation cells containing a hydrogen configuration in thermal equilibrium (see Fig.\ \ref{fig:geometry}).
The cubic simulation cells were prepared with VASP \cite{kresse1996efficient,kresse1996efficiency,kresse1999from} as described in Appendix \ref{app:vaspmd}, while the elongated simulation cells were prepared with SHRED \cite{Sharma2023,White_2022,White2020} as described in Appendix \ref{app:shredmd}.

\subsection{VASP}
\label{app:vaspmd}
The finite-temperature stopping power calculations used thermally equilibrated ionic configurations generated through separate \emph{ab initio} MD simulations.
The VASP MD simulations started from a disordered geometry where each hydrogen atom was randomly displaced from a periodic arrangement by up to \SI{0.2}{\angstrom} in each Cartesian direction.
Then, the ionic system evolved under Hellmann-Feynman forces from the electronic system and the Nos\'e-Hoover thermostat \cite{nose1984a} with a Nos\'e mass of 1 and time step of \SI{0.2}{\femto\second}.
The equilibrated geometry after \SI{1}{\pico\second} of simulated time then served as the atomic configuration for TDDFT stopping power simulations.

The MD simulations treated the electronic system with many of the same choices and parameters listed in \mbox{Sec.\ \ref{sec:vasptddft}}.
However, a reduced cutoff energy of \SI{1000}{\electronvolt} sufficed to converge the average pressure and the average free energy of the electronic system to within 3\% (0.7\%) of the \SI{2000}{\electronvolt} cutoff used in TDDFT calculations.
To further reduce computational costs, the MD simulations relaxed the electronic self-consistency criterion to energy convergence within \SI{1}{\milli\electronvolt}.

\subsection{SHRED}
\label{app:shredmd}

To generate the replicated configurations referenced in Sec.\ \ref{sec:setup}, we used \emph{ab initio} MD with the deterministic DFT capability in SHRED. 
We started with random initial ion positions in a supercell containing 256 atoms.
Their initial velocities are given by a Maxwell-Boltzmann distribution corresponding to the given temperature. 
In each MD step, SHRED started with a self-consistent-field (SCF) calculation to solve for the equilibrium electron density.
Once a converged electron density is obtained, the electronic force is derived from the Hellman-Feynman theorem, which along with ionic forces drives the motion of classical ions in each MD step. 
These SCF-MD cycles are repeated for thousands of steps to sample different ionic configurations for a real warm-dense matter system.

Specifically for the warm dense hydrogen problem considered here, we used 480 bands, the PBE exchange-correlation functional \cite{perdew1996generalized}, and the Hartwigsen-Goedeker-
Hutter (HGH) pseudopotential \cite{goedecker1996separable,hartwigsen1998relativistic} for the electron-ion interaction in SHRED SCF calculations.
The MD time step was \SI{0.1}{\femto\second} and the thermostat enforced isokinetic velocity scaling at every MD step.
Overall, the SHRED MD run gave a converged pressure of $4.82 \pm \SI{0.066}{\mega\bar}$ for the hydrogen system.
A long trajectory from such an MD run can provide multiple uncorrelated snapshots for TDDFT stopping power calculations, though this work used only a single equilibrated MD configuration for each set of TDDFT simulations.

A similar procedure was used for the fully thermalized 1024-atom configuration, but the orbital-free DFT (OF-DFT) capability in SHRED was used for this case.
The kinetic energy functional by Thomas, Fermi, and Perrot \cite{Perrot79} and the nonlinear conjugate gradient algorithm outlined by Jiang and Yang \cite{Yang04} were used to solve for the electron density.
The snapshot used for TDDFT stopping power calculations was taken after 1 ps (with an MD time step of about \SI{0.25}{\femto\second}) of OF-DFT-MD simulation time.
The supercell geometry was rectilinear with an aspect ratio of 4:1:1 and 256$\times$64$\times$64 real-space grid points (\SI{678}{\electronvolt} cutoff energy).
A pressure of $5.04 \pm \SI{0.0015}{\mega\bar}$ was calculated after discarding the first \SI{0.2}{\pico\second} for equilibration.

\bibliography{main.bib}

\end{document}